\documentclass[fleqn,usenatbib]{mnras}

\usepackage{newtxtext,newtxmath}

\usepackage[T1]{fontenc}
\usepackage{ae,aecompl}

\usepackage{graphicx}	
\usepackage{amsmath}	
\usepackage{amssymb}	
\usepackage{booktabs}
\usepackage{multirow}

\newcommand{\pcm}{\,cm$^{-2}$}	
\newcommand{\msun}{\,M$_\odot$} 
\newcommand{\mjup}{\,M$_{\rm Jup}$} 

\newcommand{\teff}{T$_{\rm eff}$}
\newcommand{\ang}{\,\AA\,}
\newcommand{\kms}{km\,s$^{-1}$}
\newcommand{\pyr}{\,yr$^{-1}$}	

\title[Signs of accretion in NLTT5306]{Signs of accretion in the white dwarf + brown dwarf binary NLTT5306}
\author[E. S. Longstaff et al.]{E. S. Longstaff$^{1}$\thanks{E-mail: el139@le.ac.uk}, S. L. Casewell$^{1}$, G. A. Wynn$^{1}$, K.L. Page$^{1}$, P. K. G. Williams$^{2}$, 
\newauthor I. Braker$^{1}$, P. F. L. Maxted$^{3}$ \\
$^{1}$Department of Physics and Astronomy, University of Leicester, University Road, Leicester LE1 7RH, UK \\
$^{2}$Harvard-Smithsonian Center for Astrophysics, 60 Garden Street, Cambridge, MA 02138, USA \\
$^{3}$Department of Physics and Astrophysics, Keele University, Keele, Staffordshire, ST5 5BG, UK \\
}

\date{Accepted XXX. Received YYY; in original form ZZZ}

\pubyear{2019}

\begin{document}
\label{firstpage}
\pagerange{\pageref{firstpage}--\pageref{lastpage}}
\maketitle

\begin{abstract}
We present new XSHOOTER spectra of NLTT5306, a 0.44 $\pm$ 0.04\msun white dwarf in a short period (101\,min) binary system with a brown dwarf companion that is likely to have previously undergone common envelope evolution. We have confirmed the presence of H$\alpha$ emission and discovered Na I absorption associated with the white dwarf. These observations are indicative of accretion. Accretion is typically evidenced by high energy emission in the UV and X-ray regime. However our \textit{Swift} observations covering the full orbital period in three wavebands (uvw1, uvm2, uvw2) revealed no UV excess or modulation. We used the X-ray non-detection to put an upper limit on the accretion rate of 2$\times$10$^{-15}$\msun yr$^{-1}$. We compare NLTT5306 to similar accreting binaries with brown dwarf donors and suggest the inferred accretion rate could be from wind accretion or accretion from a debris/dust disk. The lack of evidence for a disk implies NLTT5306 is magnetically funnelling a weak wind from a potentially low gravity brown dwarf. The upper limit on the accretion rate suggests a magnetic field as low as 0.45\,kG would be sufficient to achieve this. If confirmed this would constitute the first detection of a brown dwarf wind and could provide useful constraints on mass loss rates.
\end{abstract}




\section{Introduction}

\begin{table*}
\begin{center}
\begin{tabular}{l l c c c c c l}
\toprule
\multirow{2}{*}{Name} & \multirow{2}{*}{Type} & WD mass & BD mass & WD T$_{\rm eff}$ & BD T$_{\rm eff}$ & Period & \multirow{2}{*}{Reference} \\
 	 & 		& (\msun)   & (\msun)	& (K)		& (K) & (Minutes)	 & \\
\midrule
SDSS J121209.31 & \multirow{2}{*}{Polar} & \multirow{2}{*}{$0.8$} & \multirow{2}{*}{$-$} & \multirow{2}{*}{$9500$} & \multirow{2}{*}{$<1700$} & \multirow{2}{*}{$88.428\pm0.001$} & \citealt{burleigh06} \\
+013627.7& & & & & & & \citealt{stelzer17} \\
\midrule
\multirow{4}{*}{EF Eridani}	& \multirow{4}{*}{Polar} & \multirow{4}{*}{$>0.65$} & \multirow{4}{*}{$<0.06$} & \multirow{4}{*}{$>9500$} & \multirow{4}{*}{$<2000$} & \multirow{4}{*}{$80.64$} & \citealt{beuermann00} \\
& & & & & & & \citealt{schwope07} \\
& & & & & & & \citealt{schwope10} \\
& & & & & & & \citealt{szkody10} \\
\midrule
SDSS J125044.42 & \multirow{2}{*}{LARP} & \multirow{2}{*}{$0.60$} & \multirow{2}{*}{$0.081\pm0.005$} & \multirow{2}{*}{$10\,000$} & \multirow{2}{*}{$-$} & \multirow{2}{*}{$86$} & \citealt{steele11} \\
+154957.4& & & & & & & \citealt{breedt12} \\
\midrule
SDSS J151415.65 & \multirow{2}{*}{LARP} & \multirow{2}{*}{$\sim0.75$} & \multirow{2}{*}{$0.077$} & \multirow{2}{*}{$10\,000$} & \multirow{2}{*}{$-$} & \multirow{2}{*}{$89$} & \citealt{kulebi09} \\
+074446.5& & & & & & & \citealt{breedt12} \\
\midrule
\multirow{2}{*}{SDSS J105754.25} & \multirow{3}{*}{CV} & \multirow{3}{*}{$ 0.800\pm0.015$} & \multirow{3}{*}{$0.0436\pm0.0020$} & \multirow{3}{*}{$13300\pm1100$} & \multirow{3}{*}{$-$} & \multirow{3}{*}{$90.44\pm0.06$} & \citealt{szkody09} \\
\multirow{2}{*}{+275947.5}& & & & & & & \citealt{southworth15} \\
& & & & & & & \citealt{mcallister17} \\
\midrule
SDSS J143317.78 & \multirow{4}{*}{CV} & \multirow{4}{*}{$0.868\pm0.007$} & \multirow{4}{*}{$0.0571\pm0.0007$} & \multirow{4}{*}{$12\,800\pm200$} & \multirow{4}{*}{$2000\pm200$} & \multirow{4}{*}{$78.11$} & \citealt{littlefair08} \\
+101123.3& & & & & & & \citealt{savoury11} \\
& & & & & & & \citealt{littlefair13} \\
& & & & & & & \citealt{hern16} \\
\midrule
\multirow{5}{*}{WZ Sagittae}			 & \multirow{5}{*}{CV} & \multirow{5}{*}{$0.85\pm0.04$} & \multirow{5}{*}{$0.078\pm0.006$} & \multirow{5}{*}{$14\,800$} & \multirow{5}{*}{$\sim1900$} & \multirow{5}{*}{$81.8\pm0.90$} & \citealt{cheng97}  \\
& & & & & & & \citealt{ciardi98} \\
& & & & & & & \citealt{patterson98} \\
& & & & & & & \citealt{steeghs07} \\
& & & & & & & \citealt{harrison16} \\
\midrule
SDSS J103532.98 & \multirow{2}{*}{CV} & \multirow{2}{*}{$0.94\pm0.01$} & \multirow{2}{*}{$0.052\pm0.002$} & \multirow{2}{*}{$10100\pm200$} & \multirow{2}{*}{$ - $} & \multirow{2}{*}{$82.0896\pm0.0003$} & \multirow{2}{*}{\citealt{littlefair06}} \\
+144555.8& & & & & & & \\
\midrule
SDSS J150722.30 & \multirow{2}{*}{CV} & \multirow{2}{*}{$0.90\pm0.01$} & \multirow{2}{*}{$0.056\pm0.001$} & \multirow{2}{*}{$11\,000\pm500$} & \multirow{2}{*}{$<2450$} & \multirow{2}{*}{$66.61$} & \multirow{2}{*}{\citealt{littlefair07}} \\
+523039.8& & & & & & & \\
\bottomrule
\end{tabular}
\end{center}
\caption{A summary of the current known polars, low accretion rate polars (LARPs), and CVs with brown dwarf donors.}
\label{tab:intro}
\end{table*}

Brown dwarf companions to main sequence stars are rarely detected in orbits closer than 3\,AU \citep{metchev06}. This is thought to be due to the difficulty in forming a binary with such low mass ratios (q$\sim$0.02--0.1) \citep{gret&line06}. Consequently white dwarf -- brown dwarf binaries are rare: \citet{steele11} estimate only 0.5\% of white dwarfs have a brown dwarf companion. This scarcity exists in both detached and non-detached binaries. 

Their formation begins when the white dwarf progenitor evolves off the main sequence and expands up the giant or asymptotic branch. This expansion leads to Roche lobe overflow (RLOF), which causes the companion to become engulfed and the subsequent brief phase of binary evolution to take place in a common envelope (CE). The companion loses orbital angular momentum to the envelope, which is expelled from the system, and the companion spirals in towards the white dwarf progenitor core. Complete ejection of the envelope leaves a close white dwarf -- brown dwarf binary \citep{politano04}. At this point the two bodies (not necessarily interacting) are known as a post common envelope binary (PCEB). Currently, there are nine known detached PCEBs containing a brown dwarf companion: GD1400 (WD+L6, P=9.98\,hrs; \citealt{farihi04, dobbie05, burleigh11}), WD0137-349 (WD+L6-L8, P=116\,min; \citealt{maxted06, burleigh06, casewell15}), WD0837+185 (WD+T8, P=4.2hrs; \citealt{casewell12}), SDSS J141126.20+200911.1 (WD+T5, P=121.73\,min; \citealt{beuermann13, littlefair14}), SDSS J155720.77+091624.6 (WD+L3-5, P=2.27\,hrs; \citealt{farihi17}), SDSS J120515.80-024222.6 (WD+L0, P=71.2\,mins; \citealt{parsons17}), SDSSJ123127.14+004132.9 (WD+BD, P=72.5\,mins; \citealt{parsons17}) and EPIC212235321 (WD+BD, P=68.2\,mins; \citealt{casewell18}).

Eventually a detached PCEB may begin to interact via the loss of angular momentum from the companion's orbit (by means of e.g.\ magnetic braking or gravitational radiation), causing it to fill or come close to filling its Roche Lobe. At this point, mass transfer from the companion (also known as the donor or secondary star) onto the white dwarf will establish the system as a zero age cataclysmic variable (CV). If the white dwarf is magnetic, mass may be funnelled along the field lines onto the magnetic poles of the white dwarf. White dwarfs with magnetic fields strong enough to funnel mass along fieldlines through most of the distance between the binary components, preventing the formation of an accretion disk, are known as 'polars' \citep{ferrario15}. In the case of white dwarfs with less intense magnetic fields, the mass flow forms an accretion disk, which acts to redistribute angular momentum within the gas and facilitate accretion onto the white dwarf surface. The detailed evolution of CVs has been discussed extensively by e.g. \citealt{rappaport83,hellier01, knigge11, kalomeni16}.


The majority of CVs have stellar donors, but there are a few with sub-stellar donors and these are listed in Table~\ref{tab:intro}. Of these nine interacting systems, six have a direct detection of the secondary, either from emission lines such as those seen in the detached PCEB WD0137-349AB \citep{longstaff17} or via spectra of the brown dwarf (e.g. \citealt{hern16}).

NLTT5306 was identified by \citet{girven11} and \citet{steele11} who were searching for infrared excesses indicative of sub-stellar companions. Both \citet{girven11} and \citet{steele11} estimated the brown dwarf to be of spectral type dL5 with a minimum mass of 58$\pm$2\,M$_{\rm Jup}$ using the \textsc{dusty} atmospheric models (\citealt{chabrier00, baraffe02}). Follow up spectroscopy of this system by \citet{steele13} revealed NLTT5306 to be the shortest period white dwarf - brown dwarf binary system (P = 101.88 $\pm$ 0.02\,mins) known at the time, (although it has now been superseded by SDSS J1205-0242, SDSS J1231+0041, and EPIC212235321: \citealt{parsons17,casewell18}). 

\citet{steele13} found NLTT5306 to have a white dwarf mass of 0.44 $\pm$ 0.04\,M$_\odot$ and effective temperature (\teff) of 7756 $\pm$ 35\,K. They estimated the brown dwarf spectral type to be dL4 -- dL7 with  \teff\,$\approx$ 1700\,K and a minimum mass of 56$\pm$3M$_{\rm Jup}$. Observations of an H$\alpha$ emission feature with a radial velocity consistent with that of the Balmer lines from white dwarf led \citet{steele13} to suggest NLTT5306 is interacting, because the white dwarf is not hot enough to cause the emission via irradiation.

This paper presents spectra from XSHOOTER and photometry from The Neil Gehrels Swift Observatory (hereafter, \textit{Swift}). We also obtained radio observations using the Karl G. Jansky Very Large Array (VLA). Data reduction is described in Section~\ref{sec:obs} and the results are presented in Section~\ref{sec:results}. In Section~\ref{sec:analysis} we estimate an upper limit on the mass accretion rate and in Section~\ref{sec:discussion} we compare and contrast NLTT5306 with other weakly interacting systems. Finally, we present our conclusions in Section~\ref{sec:conclusion}.

\begin{table*}
\begin{center}
\begin{tabular}{l c c c c c}
\toprule
Filter & Target ID &	Start time &		Stop time &				Obs ID & 	UVOT exposure \\
\midrule
All&	34661&		2016-08-09T16:06:58&	2016-08-09T17:13:52&	00034661001&	1051.42 \\
\midrule
&		93147&		2017-06-07T00:52:57&	2017-06-07T01:46:57&	00093147001&	1618.32 \\
uvw1&	93147&		2017-06-07T08:50:57&	2017-06-07T09:44:01&	00093147002&	1674.38 \\
&		93147&		2017-06-07T16:52:57&	2017-06-07T17:46:07&	00093147003&	1667.89 \\
\midrule
&		93147&		2017-06-14T00:17:57&	2017-06-14T01:11:40&	00093147004&	1635.64 \\
uvm2&	93147&		2017-06-14T08:15:57&	2017-06-14T09:09:53&	00093147005&	242.206 \\
&		93147&		2017-06-14T16:15:57&	2017-06-14T17:09:41&	00093147006&	1634.65 \\
&     	34661&		2017-07-24T16:07:57&	2017-07-24T17:01:41&	00034661002&	1634.45$^1$ \\
\midrule
&		93147&		2017-06-17T00:02:57&	2017-06-17T00:57:52&	00093147007&	1562.78 \\
uvw2&	93147&		2017-06-17T08:30:57&	2017-06-17T12:06:43&	00093147008&	1501.51$^2$ \\
&		93147&		2017-06-17T17:36:57&	2017-06-17T18:30:30&	00093147009&	1645.68 \\
&		93147&		2017-06-29T18:12:57&	2017-06-29T19:48:20&	00093147010&	1634.27 \\
\bottomrule
\end{tabular}
\end{center}
\caption{Details of all \textit{Swift} observations of NLTT5306 used in this work.}
\label{table:swiftobs}
\end{table*}

\section{Observations and data reduction} \label{sec:obs}

\subsection{XSHOOTER}
\noindent We obtained 20 spectra using the XSHOOTER \citep{xshooterref} instrument mounted at the ESO VLT-UT3 (`Melipal') telescope in Paranal, Chile. The observations took place on the night of 2014-August-29 between 04:35 -- 07:12 universal time as part of programme 093.C-0211(A). We observed in the wavelength range 0.3 -- 2.48\,$\micron$ using the three independent arms: UVB, VIS, and NIR. The 20 exposures, each 300\,s long, were taken covering 1.3 orbits giving us $\sim$15 spectra per full orbit. There was a high wind of 13\,ms$^{-1}$ at the start of the night which later slowed to 8\,ms$^{-1}$. The airmass ranged between 1.02 -- 1.58 and the seeing between 0.63 -- 1.1\,arcsec. The data and standards were taken in nodding mode and were reduced using \textsc{gasgano} (v2.4.3) \citep{gasgano} following ESO's XSHOOTER pipeline (v2.5.2).

The XSHOOTER data were normalised and phase-binned using Tom Marsh's \textsc{molly} software\footnote{http://hea-www.cfa.harvard.edu/\~saku/molly/HELP.html}. The H$\alpha$ absorption feature has a narrow core and broad wings. To fit this line we followed the method from \citet{longstaff17} and used three Gaussian profiles: two for the wings and one for the core. The best fitting widths found were 30.0, 10.0, and 4.6\,nm with depths of -0.12, -0.15, and -0.24\,mJy, respectively. A fourth Gaussian was added to the previous three to model the H$\alpha$ emission feature. Its width was fixed to 0.1\,nm and its height and position were free to roam thus we were able to verify \citet{steele13} detection of H$\alpha$ emission. We searched the 0.3 -- 2.4\,$\micron$ wavelength range for features similar to those seen in \citet{longstaff17} and identified Na I absorption lines at 5889 and 5895\ang. We fit the Na I absorption lines with a single Gaussian as these lines are much narrower.

\begin{figure*}
\includegraphics[height=8.0cm]{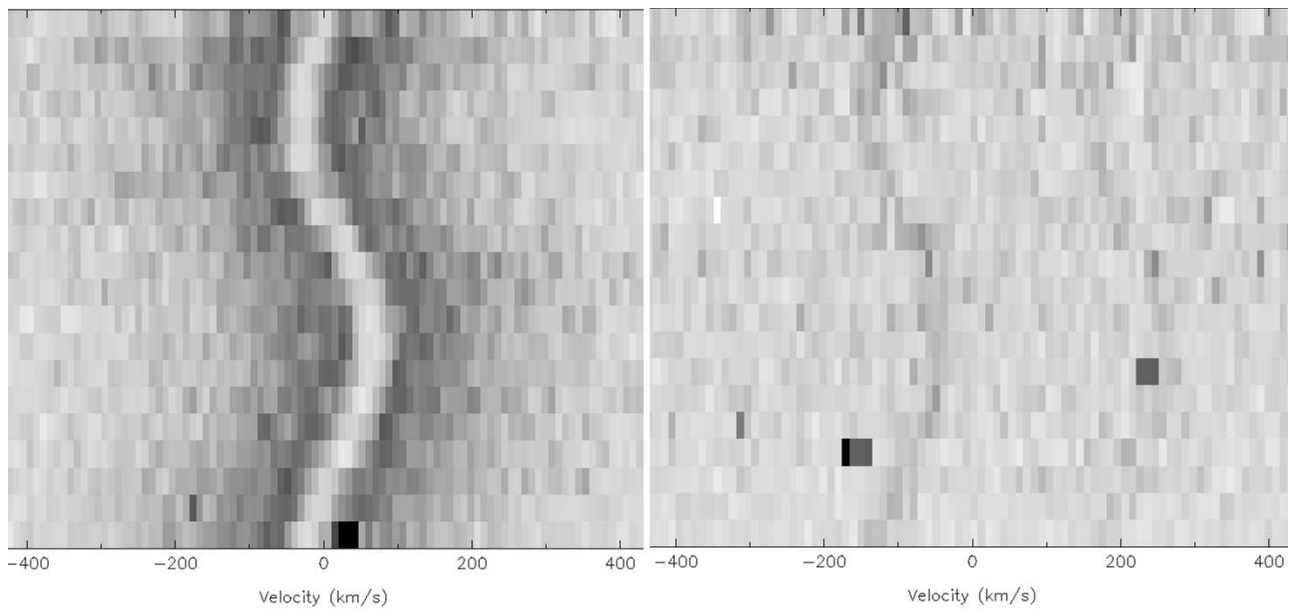}
\caption{\textbf{Left:} The trailed spectrum of the H$\alpha$ line. The emission (white) traces the path of the absorption (dark grey). \textbf{Right:} The trailed spectrum of the 5889.95\ang and 5895.92\ang Na I lines.}
\label{fig:trails}
\end{figure*}

\subsection{Swift UVOT and XRT}
In order to determine if there was any high energy emission associated with accretion from NLTT5306, we obtained a \textit{Swift} Target of Opportunity (ToO) observation (target ID:34661). We observed on 2016-August-09 in Image mode for 400\,s in each waveband. We detected the sources and achieved a signal-to-noise ratio of $>$15 in all filters. We used the standard UVOT analysis pipeline\footnote{http://www.swift.ac.uk/analysis/uvot/index.php} to derive magnitudes. For each image a source region of 5'' was selected around the source and a 20'' background region was selected from a nearby blank region. We set the background threshold to 3$\sigma$. When observing with UVOT, data are simultaneously taken by the X-ray telescope (XRT); again we followed the standard reduction pipeline to reduce these data, however, the source was not detected.

We followed up our initial ToO with further observations in the same filters as part of the GI program (target ID: 93147) to get full orbital coverage. Ideally we would have had continuous observations however \textit{Swift} has a maximum snapshot length of $\sim$1.8\,ks. Consequently our observations consisted of nine $\sim$1.5\,ks snapshots (three per filter) culminating in a total observation time of 13.1\,ks (Target ID: 00093147001 - 9). Two of the nine observations were repeated; one due to an interruption (00093147005), and one due to the observation being split into smaller snapshots significantly lowering the signal-to-noise (00093147008). All observations were taken in Event mode. 

Event mode provides a 2D sky map similar to Image Mode however each individual photon or `event' is tagged with an associated start and stop time. The reduction process is as follows. First the raw coordinate positions are converted into detector and sky coordinates using the generic \textsc{ftool} command, \textsc{coordinator}. The events are then screened using time tagged quality information to remove hot pixels using \textsc{uvotscreen}. Source and background regions are selected -- in an identical way to the Image Mode reduction -- and the background subtracted. Finally a lightcurve was created using \textsc{uvotevtlc}. This allows the user to select the time bin sizes. Each $\sim$1.5\,ks snapshot was divided into 400\,s bins for a direct comparison to our ToO observation. All observations are listed in Table~\ref{table:swiftobs}.

\begin{figure}
\rotatebox{270}{\includegraphics[height=16cm]{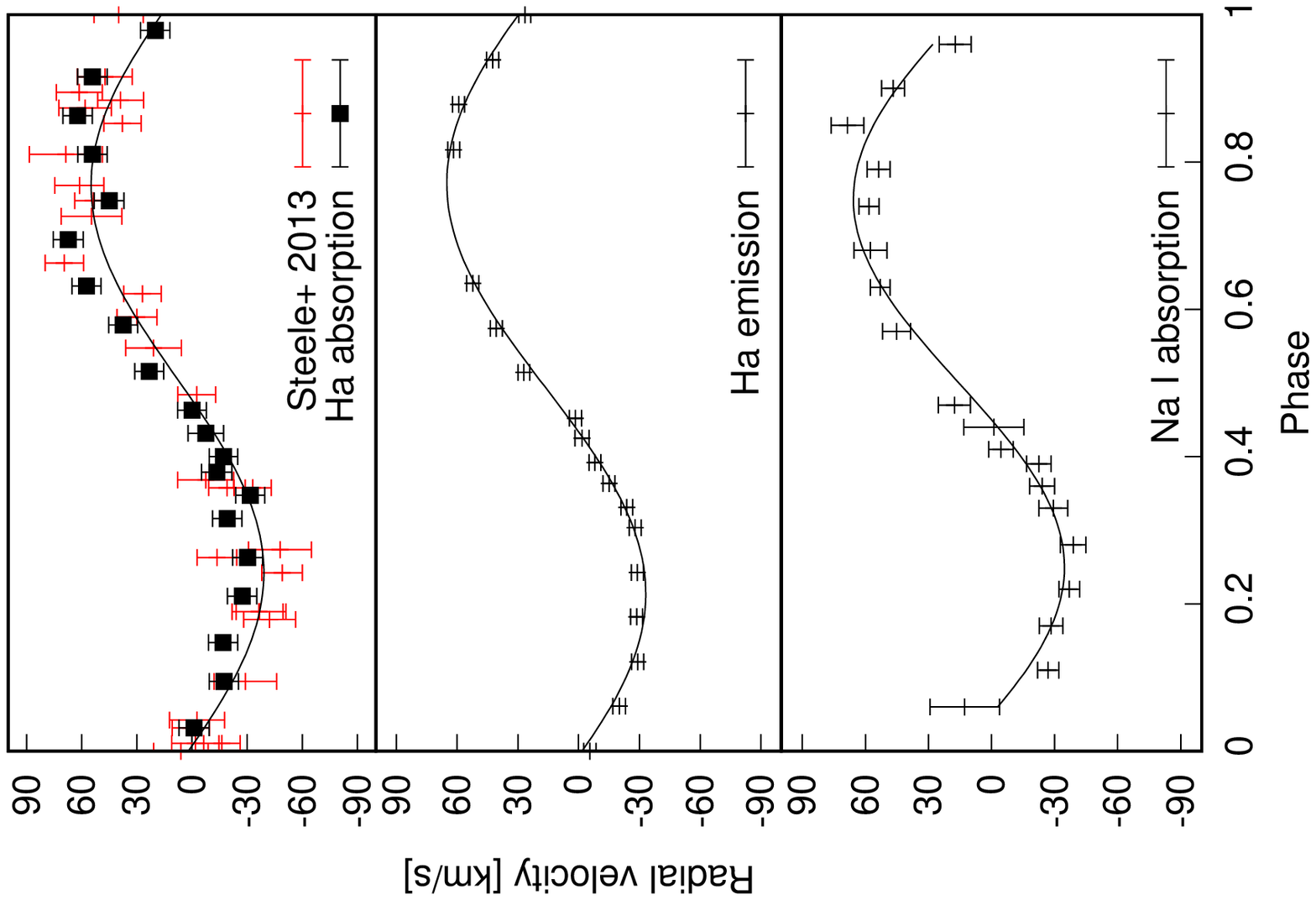}}
\caption{Radial velocity curves of the H$\alpha$ absorption, H$\alpha$ emission, and Na I absorption lines present in the NLTT5306 atmosphere.}
\label{fig:rvplot}
\end{figure}

\subsection{VLA radio observations}
We observed NLTT5306AB on 2014-May-26 with the Karl G.\ Jansky Very Large Array (VLA) at a mean frequency of 6.05~GHz using 2~GHz of bandwidth and the 8-bit samplers as in our previous work (programme ID VLA/14A-239). We observed for 3~hours to ensure we covered a full orbital period of the system.  This long observing strategy was selected as radio-emitting brown dwarfs often only emit at a narrow range of phases \citep[e.g.,][]{hallinan06}.

We calibrated and imaged the data using standard procedures and a software pipeline built on the \textsc{casa} package \citep{mcmullin07}. The image is of high quality, having a 3$\sigma$ rms noise of 6\,$\mu$Jy. We do not detect NLTT5306AB, with the image noise corresponding to an upper limit on spectral luminosity of $L_\nu < 10^{13.6}$ erg\,s$^{-1}$\,Hz$^{-1}$ at our adopted distance. To search for short-timescale radio bursts that may not have been detectable in the deep image, we additionally performed a time-domain analysis of the visibility data using the technique described in \citet{williams.2013}. We found no evidence of bursts in the time series analysis.

For comparison, \citet{coppejans16} observed U~Gem, a dwarf nova CV (WD+dM6), with the VLA and detected substantially brighter emission in outburst. In one 70-minute observation they detected a radio peak of $27.0 \pm 7.7$\,$\mu$Jy (at 8--12~GHz) that subsequently faded below their detection limit of $3\sigma = 17.5$\,$\mu$Jy; these correspond to spectral luminosities of $10^{14.45}$ and $10^{14.26}$\,erg\,s$^{-1}$\,Hz$^{-1}$, respectively, at that object's \textit{Gaia}~DR2 distance of 93.4~pc \citep[source Gaia DR2 674214551557961984;][]{gaia18a}.

\section{Results} \label{sec:results}
\subsection{XSHOOTER}

Figure~\ref{fig:trails} shows the unphased trailed spectra of the H$\alpha$ line and the Na I doublet; demonstrating that all of the lines are moving in phase with one another. We used the measured radial velocities of the H$\alpha$ absorption line from our new XSHOOTER data to refine the ephemeris of NLTT5306. We fit the equation,

\begin{equation}
V_r = \gamma + K {\rm sin}(2\pi\dfrac{t - T_0}{P}),
\label{eq:rv}
\end{equation}

\noindent to the measured radial velocities using a least-squares method allowing the systemic velocity, $\gamma$, the semi-amplitude, $K$, the zero point of the ephemeris, $T_0$ defined as when the white dwarf is furthest from the observer, and the period, $P$, to roam. Any points more than 3$\sigma$ away from the model were excluded to improve the fit. 1\kms was repeatedly added in quadrature to the standard statistical errors until a reduced $\chi^2$ of 1 was achieved resulting in $\sim$6\kms added in quadrature to the radial velocity of the H$\alpha$ absorption. When fitting the H$\alpha$ emission and Na I absorption $T_0$ and $P$ were held constant. This resulted in the errors on all radial velocity points being increased by 2\,\kms for the H$\alpha$ emission and by 0.6\,\kms\,for Na I. The error bars in Figure~\ref{fig:rvplot} reflect these error values. The best fitting orbital parameters can be found in Table~\ref{table:sysprop} and were used to phase-fold the \citet{steele13} data onto the same ephemeris.

This system is not eclipsing, however we have redefined the zero point of the ephemeris to when the unseen brown dwarf is closest to the observer. This was done to match with convention rather than \citet{steele13} initial estimate of $T_0$.

\begin{table}		
\begin{center}
\begin{tabular}{l  c  r  l}
\toprule
\multicolumn{4}{c}{System parameters} \\
\midrule
P 				& & $ 0.07075025(2)$	& Days\\
T$_0$ 			& & $ 3740.1778(8)$		& HJD-2450000\\
K$_{\rm H_{abs}}$		& & $ -46.8\pm2.5$		& \kms\\
K$_{\rm H_{emis}}$ 	& & $ -49.1\pm1.1$		& \kms\\
K$_{\rm Na}$ 		& & $ -50.2\pm6.7$		& \kms\\
$\gamma_{\rm H_{abs}}$ 		& & $ 15.6\pm1.8$		& \kms\\
$\gamma_{\rm H_{emis}}$ 		& & $ 16.0\pm1.0$		& \kms\\
$\gamma_{\rm Na}$ 			& & $ 15.5\pm3.5$		& \kms\\
\midrule
M$_{\textnormal{WD}}$  	& & 0.44 $\pm$ 0.04& M$_\odot$\\
M$_{\textnormal{BD}}$  	& & 56 $\pm$ 3&\mjup\\
R$_{\textnormal{WD}}$  	& & 0.0156 $\pm$ 0.0016 & R$_\odot$\\
Separation, a 			& & 0.566 $\pm$ 0.005& R$_\odot$\\
Mass ratio, q 			& &0.12 $\pm$ 0.01& --\\
\bottomrule
\end{tabular}
\end{center}
\caption{The parameters of the orbit of NLTT5306. K$_1$ and $\gamma_1$ found by fitting a combined data set of new XSHOOTER and \citet{steele13} radial velocities of the H$\alpha$ absorption line. K$_2$ and $\gamma_2$ found by fitting our H$\alpha$ emission and Na I absorption lines. The values below the centre line are taken from \citet{steele13} and listed here for convenience due to their use in this work.}
\label{table:sysprop}
\end{table}

\subsection{Swift} \label{sec:swiftres}
Prior to our observations we predicted magnitudes in each of the \textit{Swift} filters. We convolved a pure Hydrogen, homogeneous, plane-parallel, LTE white dwarf model (\teff\,= 7750\,K; log g = 7.75, \citealt{koester10}) with the \textit{Swift} filter profiles: uvw1, uvm2, and uvw2. Our results can be found in Table~\ref{tab:mags}.

\begin{figure}
\rotatebox{270}{\includegraphics[height=8cm]{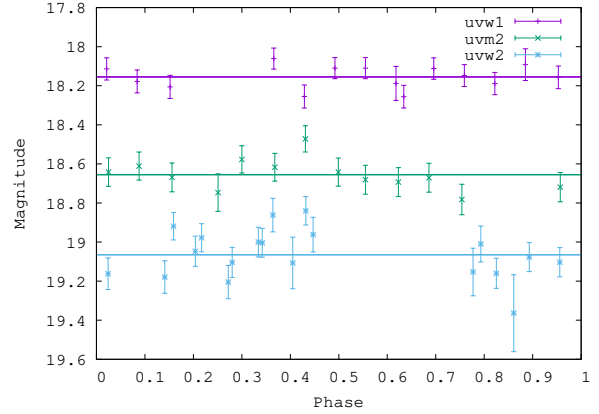}}
\caption{Measured magnitudes in the filters: uvw1, uvm2, and uvw2. Observations were taken covering as much of an orbit as possible. The horizontal lines are the average of the data points in each filter.}
\label{fig:mags}
\end{figure}

The combination of our ToO and GI observations allowed us to create a lightcurve of the system for a full orbital period. Figure~\ref{fig:mags} shows we did not detect any significant periodic variability on the period of the system. Plotting a straight line at the average magnitude for each waveband shows how little variability there is. We calculate the median average deviation as a percentage to be 1 per cent. To investigate this small scatter we compared the lightcurves of our target to those of three other sources in the field. These reference sources showed similar percentage deviations and the scatter in all three filters is consistent with Gaussian statistics, thus we conclude there is no detected variation in our data. 

\begin{table}
\begin{center}
\begin{tabular}{l l c c}
\toprule
Filter 	& Effective  	& Predicted 	& Measured 	\\
 		& wavelength 	& magnitude 	& magnitude \\
 		& (\AA) 		& & \\
\midrule
uvw1		& 2589.1		& 18.45 $\pm$ 0.2		& 18.15 $\pm$ 0.1\\
uvm2		& 2228.1		& 18.94 $\pm$ 0.2 		& 18.68 $\pm$ 0.1\\
uvw2		& 2030.5		& 19.58 $\pm$ 0.2 		& 19.04 $\pm$ 0.1\\
\bottomrule
\end{tabular}
\caption{The predicted magnitudes compared to the measured magnitudes from observations using UVOT in three wavebands. The effective wavelengths have been taken from \citet{SVOfilter}.}
\label{tab:mags}
\end{center}
\end{table}

\subsection{Very Large Array}

Our continuum radio non-detection suggests that NLTT5306B is not one of the radio-active brown dwarfs, which generally have spectral luminosities greater than $10^{13.5}$\,erg\,s$^{-1}$\,Hz$^{-1}$ and short-timescale (tens of minutes) auroral radio bursts an order of magnitude brighter than this \citep{williams14}. We cannot rule out this form of activity, however. For instance, the well-studied radio-active L3.5 brown dwarf 2MASS~J$00361617{+}1821104$ \citep{berger05}, which is an order of magnitude closer to Earth than the NLTT5306AB system, has a typical radio spectral luminosity of $10^{13.4}$\,erg\,s$^{-1}$\,Hz$^{-1}$.  Unlike the known radio-active brown dwarfs NLTT5306B benefits from a substantial additional energy source in the form of its white dwarf companion. Therefore the lack of substantial radio emission from NLTT5306B may indicate that a ``clean'', stable magnetosphere is needed for brown dwarf radio emission.

\section{Analysis} \label{sec:analysis}

\subsection{H$\alpha$ emission line strength}

\begin{figure}
\rotatebox{270}{\includegraphics[height=9cm]{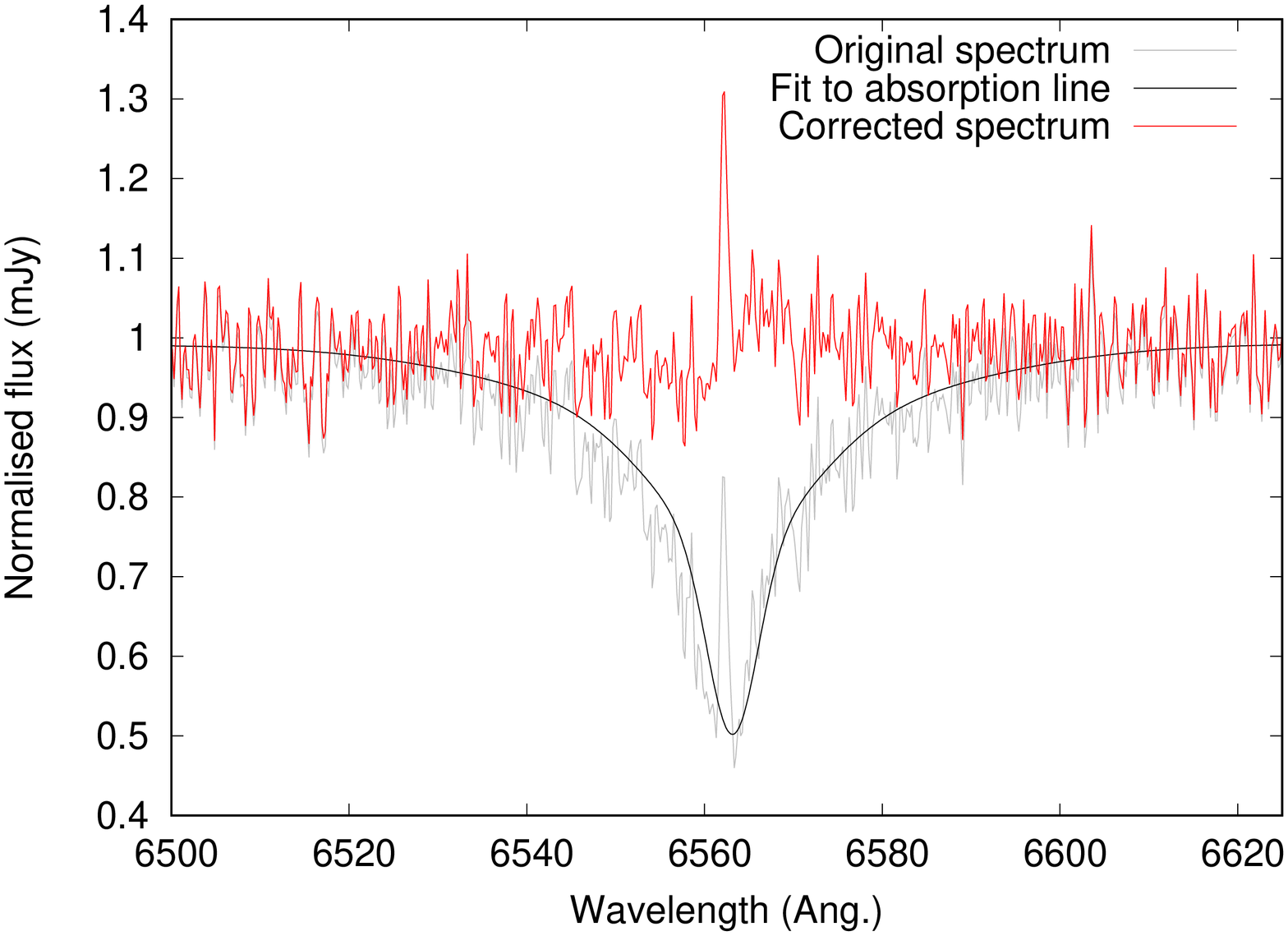}}
\rotatebox{270}{\includegraphics[height=9cm]{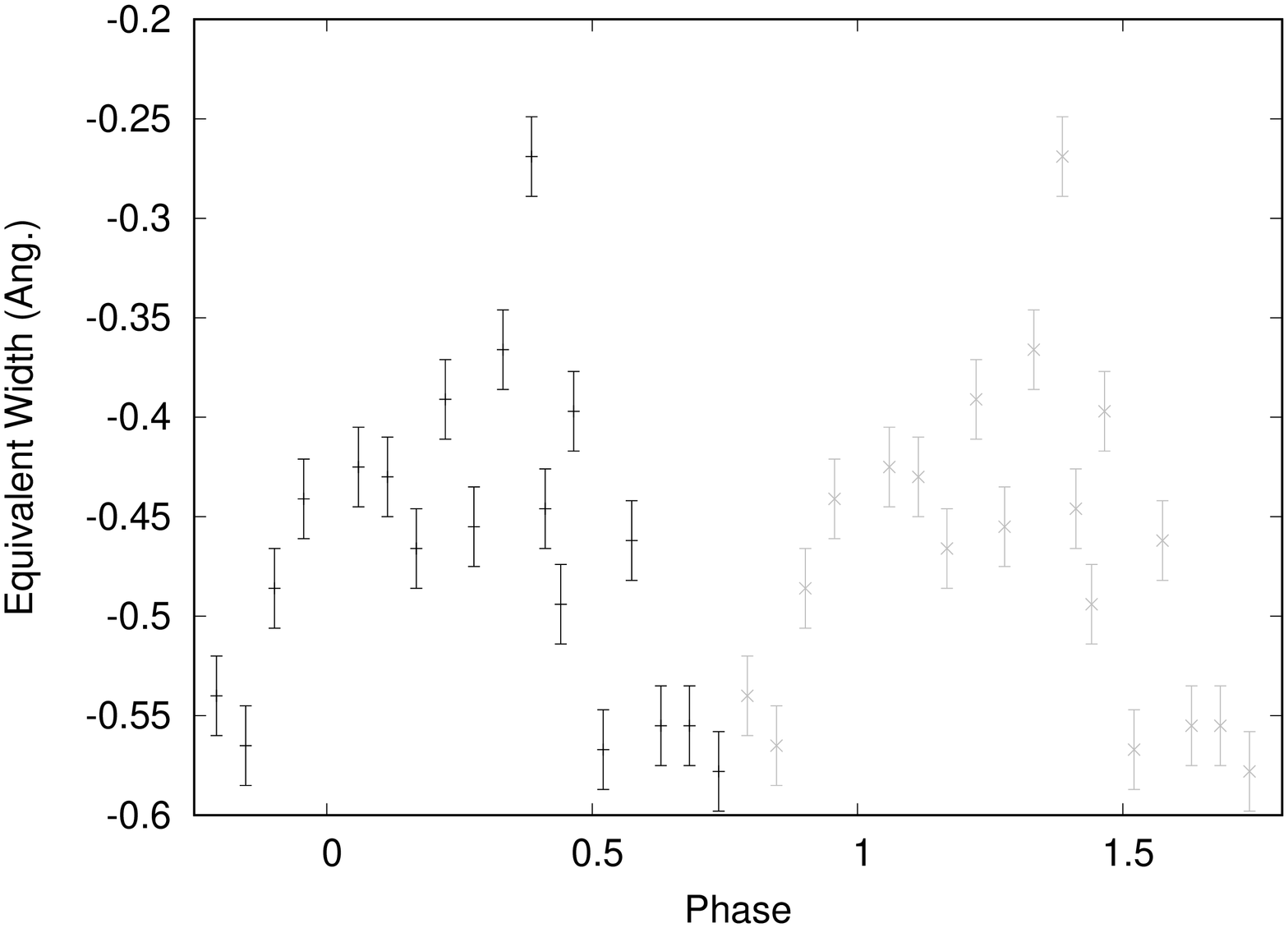}}
\caption{The upper panel shows the original spectrum (grey) with the fit to the absorption feature (black) that was subtracted to leave only the emission feature (red). The lower panel shows how the H$\alpha$ measured equivalent widths vary with phase. This has been duplicated over two phases with the duplicate data plotted in grey.}
\label{fig:EWs}
\end{figure}

To examine possible variability of the H$\alpha$ emission, we calculated the equivalent width (EW) of the line for each of the 20 XSHOOTER spectra. The equivalent width was determined using,

\begin{equation}
W = \int_{\lambda_1}^{\lambda_2} \dfrac{F_c - F_{\lambda}}{F_c} d\lambda,
\end{equation}

\noindent where $F_c$ is the continuum intensity and $F_{\lambda}$ is the intensity over the wavelength range of interest, $d\lambda$. The H$\alpha$ line is strong enough for this to be performed over a full orbit, even though the emission is in the core of the the H$\alpha$ absorption feature. To account for this we modelled and subtracted the absorption feature leaving only the emission, as shown in Figure~\ref{fig:EWs}. From the bottom panel of Figure~\ref{fig:EWs} it can be seen that the EW never reaches zero suggesting there is emission from a surface region of the white dwarf that is never fully occulted; this is consistent with Figure~\ref{fig:trails}. What also stands out in Figure~\ref{fig:EWs} is that the H$\alpha$ emission line is weakest at $\phi \sim$ 0.25 and strongest at $\phi \sim$ 0.75, i.e. there is a $\sim$0.25 phase offset between the period of the emission region and the orbital period of the system. This phase-dependent variation in line strength may give insight into the geometry of this non-eclipsing system.

\subsection{Na I absorption lines}
The only other spectroscopic lines we detect in this system are the Na I doublet pair at 5889 and 5895\ang in absorption. The signal-to-noise of these lines was too low to perform the EW analysis across a whole phase. Instead we shifted all spectra into the white dwarf rest frame and combined them. This allowed us to measure the EW of the Na I $\lambda$5889 line to be 0.12$\pm$0.02\ang and the weaker Na I $\lambda$5895 line to be 0.02$\pm$0.02\ang. Unsurprisingly the signal-to-noise of the latter line is far lower than the former. What is encouraging is the systemic velocity of the H$\alpha$ lines and the Na I absorption are in agreement with each other.

We searched for and did not detect the more commonly seen Ca II, H and K or Mg I lines. Isolated Na I absorption is unusual and, to our knowledge, has not been seen in other polluted white dwarfs (white dwarfs that display metal spectroscopic lines). \citet{rebassa16} published a list of white dwarfs with main sequence companions and lists 24 systems with both H$\alpha$ emission and Na I absorption and a further 15 systems with Na I alone. However, the Na I they observe is the 8183 and 8194\ang doublet, not the shorter wavelength line seen here, which was not within the wavelength range \citet{rebassa16} studied.

\subsection{Modelling the spectra and photometry}
\begin{figure*}
\begin{center}
\rotatebox{270}{\includegraphics[width=10cm, height=18cm]{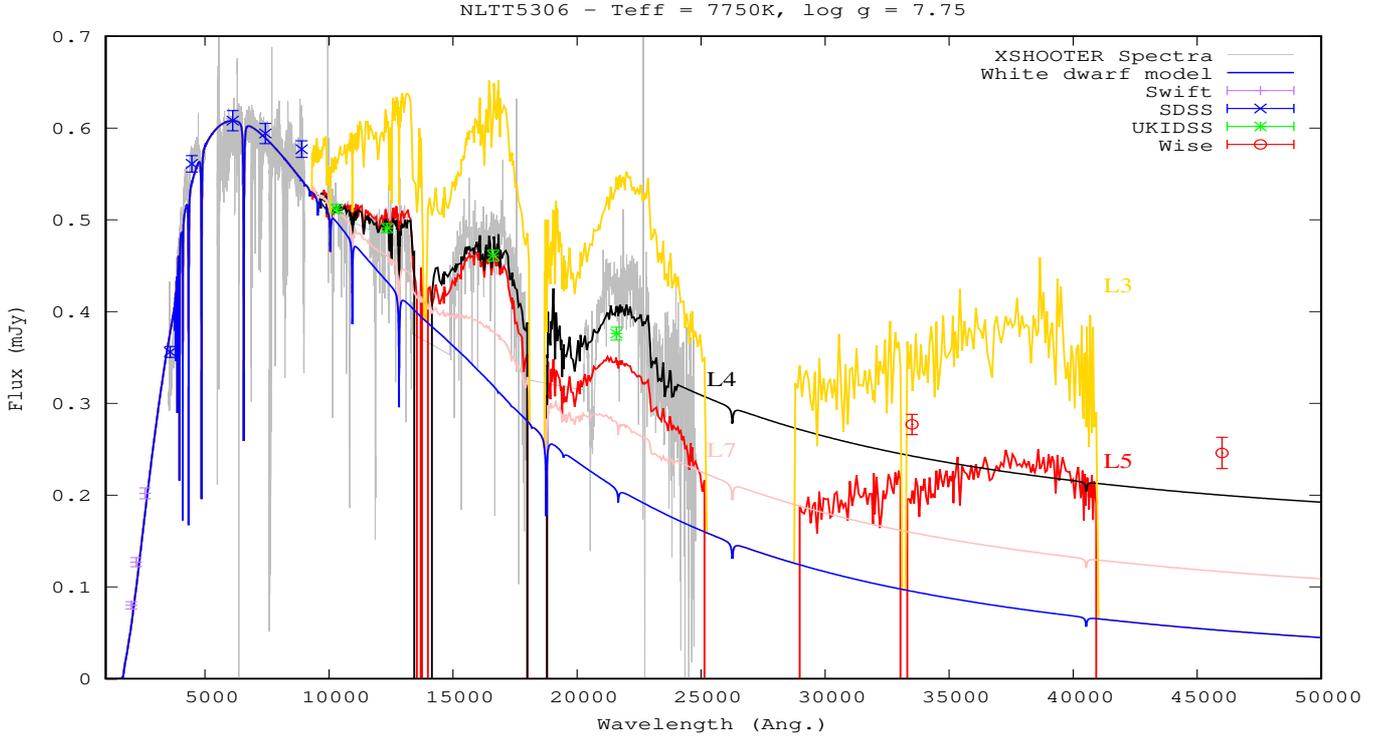}}
\caption{A combination of the white dwarf model spectrum (blue line) and a sequence of brown dwarf template spectra from L3 -- L7 (labelled). Swift photometry is plotted in purple (+), SDSS $ugriz$ in blue (x), UKIDSS $Y$, $J$, $H$ and $K$ in green (*) and Wise W1 and W2 in red (o).}
\label{fig:model_fit}
\end{center}
\end{figure*}

We combined the white dwarf model described in Section~\ref{sec:swiftres} with L3 -- L7 brown dwarf spectra \citep{cushrayvac05}. We normalised our XSHOOTER UVB, VIS, and NIR spectra to the $g$, $z$ and $J$ bands, respectively, ensuring they form a smooth continuum. From Figure~\ref{fig:model_fit} we can see that our model is consistent with the SDSS \textit{ugriz}, UKIDSS $YJHK$, and Wise (W1, W2) photometry. From Table~\ref{tab:mags} we can see our measured magnitude of the W2 band is marginally brighter than predicted; and when viewed closely the \textit{Swift} photometry sits slightly above our model in Figure~\ref{fig:model_fit}. This is likely due to an underestimate of our errors on our predicted magnitudes when convolving the white dwarf model with the \textit{Swift} filter profile.

\subsection{Placing an upper limit on the accretion rate}
There were no X-ray detection from any of the \textit{Swift} observations, placing an X-ray upper limit of 6.4$\times10^{-4}$\,counts\,s$^{-1}$ and a flux limit of 1.68$\times10^{-14}$\,ergs\,\pcm\,s$^{-1}$ using the \textit{Swift} online data analysis tools \citep{evans07, evans09} and \textit{WebPIMMS}. Using the Gaia distance to NLTT5306 of 77pc and the white dwarf mass and radius listed in Table~\ref{table:sysprop}, we find an upper limit X-ray luminosity, L$_x$ = 4$\pi$ D$^2$F, of 1$\times$10$^{28}$erg\,s$^{-1}$. From this we calculate an upper limit on the mass accretion rate to be 2$\times$10$^{-15}$\msun\pyr.

\subsection{Accretion mechanisms} \label{sec:accMech}
The main possible mechanisms for mass transfer are Roche lobe overflow or wind accretion from the donor. Typically, the mass flow forms a disc before losing angular momentum and accreting onto the surface of the main star. Alternatively, if the white dwarf is sufficiently magnetic, the material may be channelled directly along field lines down to the magnetic poles of the white dwarf.

\subsubsection{Roche lobe overflow} 
The Roche lobe is the region surrounding each object in a binary system, indicating the area of its gravitational dominance. When a star exceeds its Roche lobe, mass can be lost to the Roche lobe of its companion. This is referred to as mass transfer via Roche-lobe overflow \citep{Paczynski71, eggleton83}

Using the mass ratio of the system, $q = \frac{M_2}{M_1}$, and the orbital separation, $a$, listed in Table~\ref{table:sysprop} we can calculate an estimate of the Roche lobe radius $R_L$ of the brown dwarf. Using the \citet{breedt12} improvements on the \citet{eggleton83} calculation for the Roche lobe radius of a cataclysmic variable below the period gap in Equation 1, 
\begin{equation}
\dfrac{R_L}{a} = \dfrac{0.5126 q^{0.7388}}{0.6710 q^{0.7349} + \ln \left(1 + q^{0.3983}\right)}
\end{equation}
\noindent results in a Roche lobe radius of 0.12$\pm$0.02\,R$_\odot$. Using the \citet{steele13} value for the brown dwarf radius of 0.095$\pm$0.004\,R$_\odot$, we estimate that NLTT5306B is filling $\approx$80\% of its Roche Lobe. An important caveat to this is that the spectroscopic mass estimate from \citet{steele13} is model dependent, and was calculated by interpolating the Lyon group atmospheric models (\citealt{chabrier00, baraffe02}) given an estimated age for the white dwarf and temperature for the brown dwarf. These models have not been rigorously tested at these masses and ages so the brown dwarf may have a radius $larger$ than these models predict, and thus could be Roche lobe filling. Conversely, \citet{steele13} used the cooling age as a lower limit to the system age and so the brown dwarf may in fact be older thus potentially making the brown dwarf radius even smaller than our assumed value of 0.095\,R$_\odot$. This makes it difficult to assign robust uncertainties and to categorically say whether or not this system is Roche lobe filling.

It is possible for a companion with a predicted radius smaller than the radius of its Roche lobe to still be Roche lobe filling, if the primary star in the system is very hot, causing inflation of the radius of the brown dwarf. For example, the millisecond pulsar SAX J1808.4--3658 has a companion brown dwarf of mass 0.05\msun, which is not large enough to fill its Roche lobe. Despite this, \citet{bildsten01} detect a time averaged mass transfer rate of $\dot{M}\approx10^{-11}$\msun\pyr. This is attributed to continuous heating from the neutron star, inflating the brown dwarf to a radius of $\approx$0.13\,R$_\odot$. They estimate the day--side of their brown dwarf to be between 3800--4800\,K, much hotter than the 1700\,K \citet{steele13} predict for NLTT5306B. Similarly Kelt-1b, a 27\mjup\ brown dwarf orbiting a $\approx$6500\,K mid-F star in a 29\,hour orbit, receives a large amount of stellar insolation which suggests the radius is likely to be significantly inflated \citep{siverd12}.

Contrastingly when examining eclipsing white dwarf -- brown dwarf binary systems hotter than NLTT5306 we find the brown dwarf companions are not inflated. For example, SDSS J141126.20+200911.1 contains a 13000\,K white dwarf with a $\approx$ 50\mjup\ brown dwarf with a relatively small radius of 0.072\,R$_{\odot}$ \citep{littlefair14}. Similarly SDSS J1205-0242 has a $\sim$23600\,K white dwarf that irradiates a $\approx$ 50\mjup brown dwarf with a radius of 0.081\,R$_{\odot}$ \citep{parsons17}. Additionally,  the non-eclipsing system, WD0137-349AB, contains a 16\,500\,K white dwarf and a brown dwarf with a similar mass, yet accretion signatures are not seen. Therefore it is unlikely that NLTT5306A is hot enough to significantly inflate the brown dwarf.

Interestingly, NLTT5306B displays a distinct triangular shape in the $H$--band (see Fig. 3 in \citet{steele13}), suggestive of low surface gravity \citep{kirkpatrick06} which could mean the brown dwarf atmosphere is extended. This feature is usually only seen in young (e.g. 10s of Myr) substellar objects \citep{allers13} and NLTT5306B is estimated to be $\sim$0.5\,Gyr, and the cooling age of the white dwarf is 710\,Myr \citep{steele13}.

\subsubsection{Magnetic interaction}

The companions in Low Accretion Rate Polars (LARPs) and, possibly, some polars are thought to slightly underfill their Roche Lobes, with accretion taking place through magnetic funnelling of a wind  from (or material close to the surface of) the secondary onto the poles of the white dwarf \citep{king85}. These systems are characterised by containing white dwarfs with strong magnetic fields of $>$10\,MG \citep{schwope02}. The magnetic field is typically detected through Zeeman splitting of spectral lines and/or cyclotron humps. We do not see any cyclotron humps in our data; nor do we detect any Zeeman splitting of the H lines.

The upper limit on the accretion rate implies a critical magnetic field strength required to prevent the formation of an accretion disk and, instead, magnetically channel the mass flow. To disrupt disk formation, the magnetic radius, $r_m$, which de-marks the region in which the magnetic field of the white dwarf is expected to dominate the accretion dynamics,  would have to be at least equal to the circularisation radius, $R_{\textnormal{circ}}$, given by \citealt{frankkingraine},

\begin{equation}
\frac{R_{\textnormal{circ}}}{a} = (1 + q)(0.5 - 0.227 \textnormal{log}_{10}(q))^4 ,
\end{equation}
\noindent where $q$ is the mass ratio and $a$ is the orbital separation. For NLTT5306 we find this value to be $1.1 \times 10^{10}$\,cm ($\approx$ 10R$_{WD}$) using the component masses and separation (Table~\ref{table:sysprop}). By equating the ram pressure, $\rho v^2$, to the magnetic pressure, $B^2/8\pi$, the magnetic radius can be written as,

\begin{equation}
r_m = 5.1 \times 10^8 \dot{M}_{16}^{-\frac{2}{7}} m_1^{-\frac{1}{7}} {\mu}_{30}^{\frac{4}{7}}	 ,
\label{eq:bfield}
\end{equation}

\noindent where $\dot{M}_{16}$ is the mass accretion rate in units of 10$^{16}$ gs$^{-1}$, m$_1$ is the mass of the white dwarf in solar masses and ${\mu}_{30}$  is the white dwarf magnetic moment in units of 10$^{30}$\, Gcm$^{3}$ (see e.g.\ \citealt{frankkingraine}). Setting Equation~\ref{eq:bfield} equal to the circularisation radius and adopting the white dwarf radius in Table~\ref{table:sysprop}, we find a critical magnetic field of 0.45$\pm$0.02\,kG to prevent the formation of an accretion disc. This is a very low field limit and we would expect any white dwarf to be at least this magnetic (\citealt{scaringi17}), consistent with the lack of any observational evidence for a substantial accretion disk. The accretion flow in NLTT5306 is therefore highly likely to be magnetically controlled. 

\subsubsection{Wind from a companion}

In some cases the compact primary is able to capture a fraction of material from a companion via a wind. This is known as wind accretion and, typically, the wind originates from stellar companions. For example, V471 Tau captures the coronal mass ejections from the K-dwarf \citep{bois1991,sion12}. Similarly, LTT 560 displays H$\alpha$ emission on the white dwarf caused by stellar activity from the secondary late-type main-sequence star \citep{tappert11}. Unlike stellar objects, brown dwarfs are degenerate and much cooler than stars which is likely the reason why a brown dwarf wind has never been detected. It is unlikely that a wind from NLTT5306B is being generated through irradiation alone since we do not see anything of the sort in similar more higly irradiated systems, including those with very short periods (e.g. \citealt{casewell18}). A low rate of mass loss via a wind is possible if the low gravity implied by the $H$--band is confirmed \citep{allers13}.

\section{Discussion} 
\label{sec:discussion}
The presence of H$\alpha$ emission and Na I absorption in the atmosphere of NLTT5396A suggests accretion due to the short (days to weeks) sinking time-scale in DA white dwarfs \citep{wyatt14}. However, the lack of high energy emission places strong constraints on the possible rate of accretion. Modulation of the H$\alpha$ emission line strength suggests that the accretion region or `hot-spot' is moving with respect to the observer which could give an insight into the geometry of this non-eclipsing system. To adequately assess how NLTT5306 fits into the context of other interacting systems we compare our accretion rate upper limit to that of similar binaries.

\subsection{Accretion rate} \label{sec:accrate}
The systems in Table~\ref{tab:intro} have accretion rates orders of magnitude higher than NLTT5306. The CV WZ~Sagittae has a rate of 7.5$\times 10^{-13}$\msun\,yr$^{-1}$, and the polars SDSS J121209+013627 and EF Eri: $\sim$10$^{-13}$\msun\,yr$^{-1}$ \citep{burleigh06,koen06} and 3.2x10$^{-11}$\msun\,yr$^{-1}$ \citep{schwope07}, respectively. Pre-polars have slightly lower rates at 10$^{-14}$ - 10$^{-13}$\msun\,yr$^{-1}$ \citep{schmidt07}. White dwarfs with stellar companions that accrete via a wind have accretion rates from 10$^{-19}$ to 10$^{-16}$\msun\pyr\   \citep{debesJ06,kawka08} and the rate largely depends on the spectral type of the companion. 

We can also compare our accretion rate to those of polluted white dwarfs. There are a handful of white dwarfs accreting from gas disks, such as SDSS J122859+104032, which has an upper limit accretion rate of 5$\times\,10^{-17}$\msun\pyr (\citealt{gansicke06,brinkworth09}). The white dwarfs with dust disks have similar accretion rates, for example, $\dot{M}\sim 7\times$10$^{-17}$\msun\pyr \citep{vonhippel07};  10$^{-19}$ to 10$^{-16}$ \msun\pyr\ \citep{farihi18}. 

SDSS J155720.77+091624.6, a unique system consisting of a white dwarf - brown dwarf binary, that also has a circumbinary debris disk has a steady state of accretion at an inferred rate of $\dot{M}\simeq 9.5\times$10$^{-18}$\msun\pyr \citep{farihi17}, while WD1145+017, a system that has multiple transiting objects thought to be disintegrating planetesimals \citep{vanderberg15}, has a steady state mass accretion rate of 6.7$\times$10$^{-16}$\msun\,yr$^{-1}$ \citep{xu16}.  


One system that does have a similar accretion rate is LTT 560 \citep{tappert07}. LTT 560 is a white dwarf + M-dwarf PCEB that displays H$\alpha$ emission associated with the white dwarf. The mass accretion rate was estimated to be $\sim$~5$\times$10$^{-15}$\msun\pyr by \citet{tappert11} and the accretion mechanism is thought to be a stellar wind from the M dwarf companion.

The above comparisons put our upper limit well below that of CVs and polars, by at least two orders of magnitude, and an order of magnitude above rates due to wind and gas/dust disk accretion. From Figure~\ref{fig:model_fit} we can see the Wise photometry fits with a white dwarf + L3-5 dwarf model with no sign of an infrared excess indicative of a disk. Nor do we see any H$\alpha$ emission from anywhere other than the white dwarf surface. This evidence implies that no disk has formed therefore the mechanism for accretion is more likely to be magnetic funnelling of a light brown dwarf wind.

\section{Conclusions} \label{sec:conclusion}

H$\alpha$ emission and Na I absorption associated with the white dwarf lead us to believe accretion is happening in NLTT5306. Our lower limit on the accretion rate and the lack of any observational evidence for an accretion disk, suggests accretion may be occurring via magnetic channelling of a wind from the donor, however this could challenge the idea of NLTT5306B being a degenerate object. The phase-offset H$\alpha$ line strength variability indicates we could be seeing orbital variation of a hot-spot on the white dwarf surface, supporting the magnetically channelled wind interpretation. Yet, the lack of any detectable high energy emissions, Zeeman splitting, or cyclotron humps conflicts with this.

If we consider the possibility that the triangular shape of the H--band does indeed indicate low surface gravity, then it may be possible for a wind to escape from NLTT5306B. The low magnetic field (0.45\,kG) required to prevent a disk forming from matter accreting at a rate of 2$\times$10$^{-15}$\msun\pyr 
indicates there is scope for such a wind to be magnetically channelled through the bulk of the system, with, at the same time, the field and mass transfer rate being too low to produce detectable Zeeman or cyclotron features. It may be possible the magnetic field of the white dwarf plays a role in the stripping of any loosely bound material close to the surface of NLTT5306B. 





This work demonstrates that NLTT5306 is a useful target for mapping the transition from PCEB to pre-CV. Our broad spectrum observations clearly indicate that a higher signal-to-noise is required. For example, systems with lower accretion rates than NLTT5306 have X-ray detections (e.g. \citet{farihi18}), requiring observations using more sensitive observatories such as XMM-Newton. Furthermore, magnetic fields weaker than 2\,MG are only detectable for high S/N spectra (e.g. \citet{tout08}) therefore thorough circular spectro-polarimetry observations are required to verify a magnetic field.

\section{Acknowledgements}

We thank Tom Marsh for the use of the molly software, and Lilia Ferrario for useful conversations regarding magnetic white dwarfs. This research has made use of the SVO Filter Profile Service (http://svo2.cab.inta-csic.es/theory/fps/) supported from the Spanish MINECO through grant AyA2014-55216. This work is based on observations made with ESO telescopes at La Silla Paranal Observatory under programme ID 093.C-0211(A). This work made use of data supplied by the UK Swift Science Data
Centre at the University of Leicester. E. S. Longstaff acknowledges the support of an STFC studentship. S. L. Casewell acknowledges support from the University of Leicester Institute for Space and Earth Observation. K.L. Page acknowledges support from the UK Space Agency. This work was supported by the Science and Technology Facilities Council [ST/M001040/1]. P.K.G.W. acknowledges support for this work from the National Science Foundation through Grant AST-1614770. The National Radio Astronomy Observatory is a facility of the National Science Foundation operated under cooperative agreement by Associated Universities, Inc.




\bibliographystyle{mnras}
\bibliography{references} 


\bsp	
\label{lastpage}
\end{document}